\documentclass[final,5p,times,twocolumn]{elsarticle}

\usepackage[T1]{fontenc}
\usepackage[utf8]{inputenc}

\usepackage{amsmath}
\usepackage{amsfonts}
\usepackage{amssymb}
\usepackage{amsxtra}
\usepackage{array}
\usepackage{graphicx}
\usepackage{hepunits}
\usepackage{units}
\usepackage{color}
\usepackage{xspace}

\definecolor{purple}{rgb}{0.5,0,0.5}
\definecolor{blue}{rgb}{0.0,0,0.9}
\definecolor{prdblue}{rgb}{0.133,0.118,0.498}
\usepackage[colorlinks=true, pdfstartview=FitV, linkcolor=prdblue, citecolor= prdblue, urlcolor=prdblue]{hyperref}

\usepackage{CJKutf8}

\usepackage[mathscr,scaled=1.15]{urwchancal}
\DeclareFontFamily{OT1}{pzc}{}
\DeclareFontShape{OT1}{pzc}{m}{it}%
{<-> s * [1.15] pzcmi7t}{}
\DeclareMathAlphabet{\mathpzc}{OT1}{pzc}{m}{it}

\newcommand{\be}{\begin{equation}}
\newcommand{\bea}{\begin{eqnarray}}
\newcommand{\ee}{\end{equation}}
\newcommand{\eea}{\end{eqnarray}}

\def\1eq#1{Eq.~(\ref{#1})}

\def\2eqs#1#2{Eqs.~(\ref{#1}) and~(\ref{#2})}
\def\3eqs#1#2#3{Eqs.~(\ref{#1}),~(\ref{#2}) and~(\ref{#3})}

\biboptions{sort&compress}

\journal{Physics Letters B}

\begin{document}

\begin{CJK}{UTF8}{song}

\begin{frontmatter}

\title{$\,$\\[-7ex]\hspace*{\fill}{\normalsize{\sf\emph{Preprint no}. NJU-INP 047/21}}\\[1ex]
Pion charge radius from pion+electron elastic scattering data
}

\author[NJU,INP]{Zhu-Fang Cui} 
\ead{phycui@nju.edu.cn}

\author[ECT]{Daniele Binosi}
\ead{binosi@ectstar.eu}

\author[NJU,INP]{Craig D. Roberts\corref{cor2}}
\ead{cdroberts@nju.edu.cn}

\author[HZDR,RWTH]{Sebastian M.~Schmidt}
\ead{s.schmidt@hzdr.de}

\address[NJU]{
School of Physics, Nanjing University, Nanjing, Jiangsu 210093, China}
\address[INP]{
Institute for Nonperturbative Physics, Nanjing University, Nanjing, Jiangsu 210093, China}

\address[ECT]{
European Centre for Theoretical Studies in Nuclear Physics
and Related Areas, Villa Tambosi, Strada delle Tabarelle 286, I-38123 Villazzano (TN), Italy}

\address[HZDR]{
Helmholtz-Zentrum Dresden-Rossendorf, Dresden D-01314, Germany}

\address[RWTH]{
RWTH Aachen University, III. Physikalisches Institut B, Aachen D-52074, Germany}


\begin{abstract}
With the aim of extracting the pion charge radius, we analyse extant precise pion+electron elastic scattering data on $Q^2 \in [0.015,0.144]\,$GeV$^2$ using a method based on interpolation via continued fractions augmented by statistical sampling.  The scheme avoids any assumptions on the form of function used for the representation of data and subsequent extrapolation onto $Q^2\simeq 0$.  Combining results obtained from the two available data sets, we obtain $r_\pi = 0.640(7)\,$fm, a value $2.4\,\sigma$ below today's commonly quoted average.  The tension may be relieved by collection and similar analysis of new precise data that densely cover a domain which reaches well below $Q^2 = 0.015\,$GeV$^2$.
Considering available kaon+electron elastic scattering data sets, our analysis reveals that they contain insufficient information to extract an objective result for the charged-kaon radius, $r_K$.  New data with much improved precision, low-$Q^2$ reach and coverage are necessary before a sound result for $r_K$ can be recorded.
\end{abstract}


\end{frontmatter}

\end{CJK}

\noindent\textbf{1.$\;$Introduction}.
The pion is Nature's lightest hadron \cite{Zyla:2020zbs}, with its mass, $m_\pi$, known to a precision of $0.0001$\%.  Indeed, with a mass similar to that of the $\mu$-lepton, the pion can seem unnaturally light.  Realising this fact, contemporary theory explains the pion as simultaneously both a pseudoscalar bound-state of a light-quark and -antiquark and, in the absence of Higgs-boson couplings into quantum chromodynamics (QCD), the strong interaction sector of the Standard Model of particle physics (SM), the massless Nambu-Goldstone boson that emerges as a consequence of dynamical chiral symmetry breaking (DCSB).  This dichotomy is readily understood so long as a symmetry-preserving treatment of QCD's quantum field equations is used \cite{Lane:1974he, Politzer:1976tv, Delbourgo:1979me, Maris:1997hd, Brodsky:2012ku, Qin:2020jig}.  Following upon such proofs, it has become clear that pion observables provide the cleanest window onto the phenomenon of emergent hadron mass (EHM), \emph{viz}.\ the origin of more than 98\% of visible mass in the Universe \cite{Horn:2016rip, Aguilar:2019teb, Anderle:2021wcy, Roberts:2021nhw, Arrington:2021biu}.

The pion was discovered more than seventy years ago \cite{Lattes:1947mw}, twelve years after its existence was first predicted \cite{Yukawa:1935xg}.  Yet, in all this time, very little has been revealed empirically about its internal structure \cite{Chang:2021utv}.  This is largely because pions are unstable, e.g., charged pions decay via weak interactions with a strength modulated by the pion leptonic decay constant \cite{Zyla:2020zbs}: $f_\pi = 0.0921(8)\,$GeV.

A simple case in point is the electric radius of the charged pion.  With results in hand for the charged-pion elastic form factor, $F_\pi(Q^2)$, where $Q^2$ is the momentum-transfer-squared, then the radius-squared is obtained via:
\begin{equation}
\label{Eqpirad}
r_\pi^2 = -\frac{6}{F_\pi(0)} \left. \frac{d}{dQ^2} F_\pi(Q^2) \right|_{Q^2=0}\,,
\end{equation}
i.e., from the slope of the form factor at $Q^2=0$.  Numerous model and QCD theory calculations of $r_\pi$, $f_\pi$ are available, for both physical pions and pseudoscalar mesons built from heavier current-quarks; and it has been found that the product of radius, $r_{0^-}$, and leptonic decay constant, $f_{0^-}$, for ground-state pseudoscalar mesons is practically constant on a large domain of meson mass:
\begin{equation}
\label{Eqfpirpi}
f_{0^-} r_{0^-} \approx {\rm constant} \,|\; 0 \leq m_{0^-} \lesssim 1\,{\rm GeV},
\end{equation}
see, e.g., Refs.\,\cite[Fig.\,1]{Cloet:2008fw}, \cite[Fig.\,2A]{Chen:2018rwz}.  This prediction is a manifestation of EHM as expressed in pseudoscalar meson Bethe-Salpeter wave functions.

The most direct available empirical means to extract $r_\pi$ is measurement of $F_\pi(Q^2)$ in pion+electron $(\pi e)$ elastic scattering at low-$Q^2$ and subsequent use of Eq.\,\eqref{Eqpirad}.  Existing estimations of this radius are broadly compatible, but the precision is at least $10\,000$-times worse than that with which $m_\pi$ is known.  Given the fundamental character of the pion, e.g., its role in binding nuclei and its connections with EHM, this is unsatisfactory.  $m_\pi$ and $r_\pi$ are also correlated \cite[Fig.\,2]{Aguilar:2019teb} and a SM solution will deliver values for both; hence, precise values are necessary to set solid benchmarks for theory.

Data on low-$Q^2$ $\pi e \to \pi e$ scattering has been collected in four experiments \cite{Amendolia:1984nz, Amendolia:1986wj, Dally:1982zk, GoughEschrich:2001ji} and today's commonly quoted value of the pion radius \cite{Zyla:2020zbs}:
\begin{equation}
\label{PDFradius}
r_\pi = 0.659 \pm 0.004\,{\rm fm},
\end{equation}
is obtained from the analysis of the sets in Ref.\,\cite{Amendolia:1986wj, Dally:1982zk, GoughEschrich:2001ji},
%
%
augmented by information from analyses of $e^+  e^- \to \pi^+ \pi^-$ reactions \cite{Ananthanarayan:2017efc, Colangelo:2018mtw}.  In the past two years, this value has dropped $1.5\,\sigma$ from the value quoted earlier \cite{Tanabashi:2018oca}: $r_\pi = 0.672 \pm 0.008\,$fm.

Following the controversy surrounding the proton charge radius, which began ten years ago \cite{Pohl:2013yb, Gao:2021sml}, it is now known that a reliable extraction of radii from lepton+hadron scattering and the associated electric form factor requires precise data that are densely packed at very-low-$Q^2$, e.g., with an electron beam energy of $1.1\,$GeV, the PRad collaboration collected $33$ proton form factor data on the domain $2.1\times 10^{-4} \leq Q^2/{\rm GeV}^2\leq 0.06$ \cite{Xiong:2019umf}.  Moreover, in analysing that data, every conceivable effort must be made to eliminate all systematic error inherent in the choice of data fitting-function \cite{Kraus:2014qua, Lorenz:2014vha, Griffioen:2015hta, Higinbotham:2015rja, Hayward:2018qij, Zhou:2018bon, Alarcon:2018zbz, Higinbotham:2019jzd, Hammer:2019uab}.

Importantly, a scheme is now available that avoids any specific choice of fitting function in analysing form factor data.  Refined in an array of applications to problems in hadron physics, particularly those which demand model-independent interpolation and extrapolation, \emph{e.g}.\ Refs.\,\cite{Tripolt:2016cya, Chen:2018nsg, Binosi:2018rht, Binosi:2019ecz, Eichmann:2019dts, Yao:2020vef, Yao:2021pyf}, and typically described as the statistical Schlessinger point method (SPM) \cite{PhysRev.167.1411, Schlessinger:1966zz}, the approach produces a form-unbiased interpolation of data as the basis for a well-constrained extrapolation and, hence, radius determination \cite{Cui:2021vgm}.  It is therefore worth reconsidering available low-$Q^2$ pion form factor data with a view to employing the SPM in a fresh determination of $r_\pi$.

\medskip

\begin{figure}[t]
\centerline{%
\includegraphics[clip, width=0.45\textwidth]{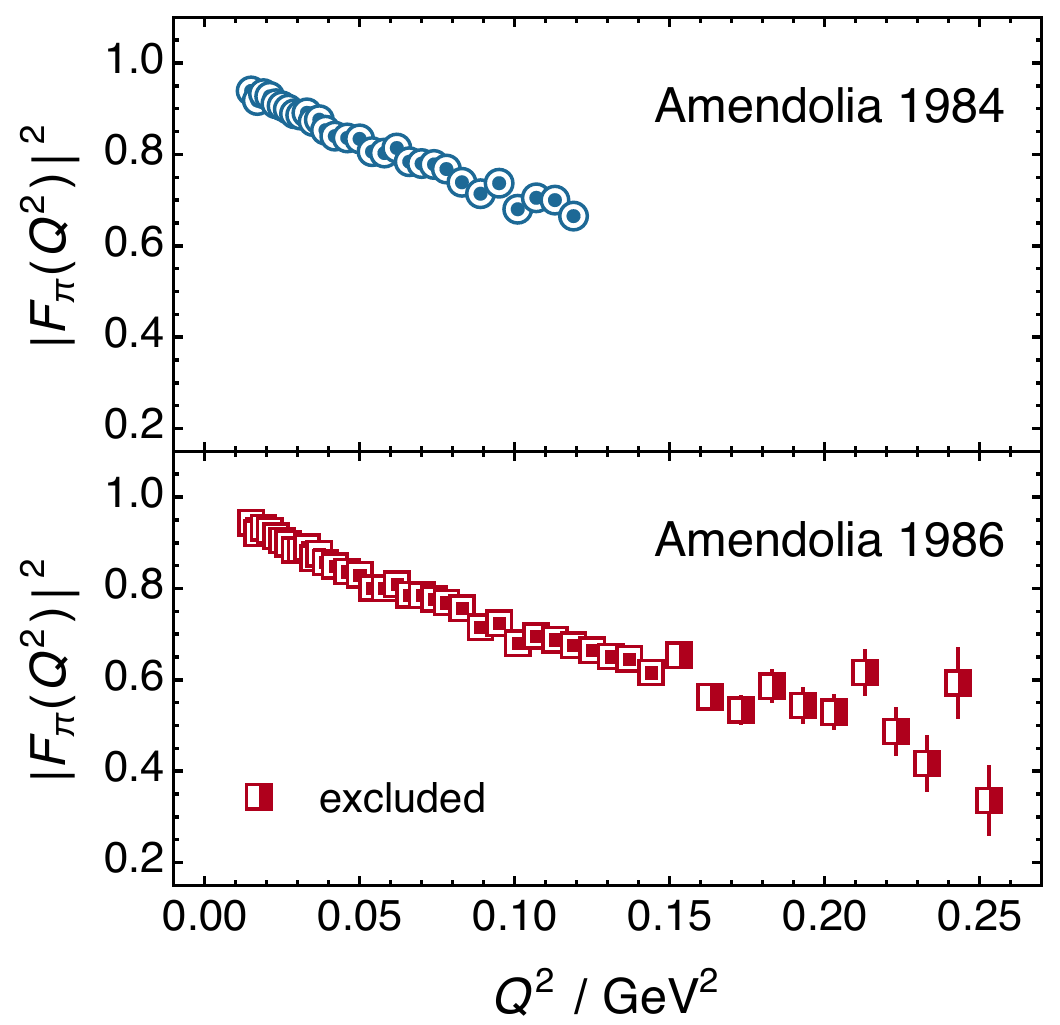}}
\caption{\label{FAmendolia}
Pion form factor data collected at CERN by the NA7 Collaboration: upper panel -- \cite{Amendolia:1984nz}; lower panel -- \cite{Amendolia:1986wj}.  Although these data were obtained thirty-five years ago, they remain the most dense and extensive sets of low-$Q^2$ data available.
Concerning the 1986 set \cite{Amendolia:1986wj}, our analysis uses the square-within-square data; the half-shaded data only introduce noise.
}
\end{figure}

\noindent\textbf{2.$\;$Pion form factor}.
The most recent $\pi e \to \pi e$ experiment was performed twenty years ago \cite{GoughEschrich:2001ji}, but the densest and most extensive sets of low-$Q^2$ data were collected roughly thirty-five years ago at CERN by the NA7 Collaboration \cite{Amendolia:1984nz, Amendolia:1986wj}.  The sets are depicted in Fig.\,\ref{FAmendolia}.
Within mutual uncertainties, the radius determination published in Ref.\,\cite{GoughEschrich:2001ji} agrees with those reported in Refs.\,\cite{Amendolia:1984nz, Amendolia:1986wj}; but referred to the results quoted in Ref.\,\cite{Amendolia:1986wj}, the central value in Ref.\,\cite{GoughEschrich:2001ji} is more than $2\sigma$ smaller.

Reviewing available low-$Q^2$ $\pi e \to  \pi e$ data, only the sets obtained by the NA7 Collaboration \cite{Amendolia:1986wj, Amendolia:1984nz} are worth considering for reanalysis using the SPM:
the $Q^2$-coverage, density, and precision of the data in Ref.\,\cite{Dally:1982zk} are too poor to contribute anything beyond what is already available in the NA7 sets;
and Ref.\,\cite{GoughEschrich:2001ji} does not provide pion form factor data, reporting a radius instead by assuming a monopole form factor.

\medskip

\noindent\textbf{3.$\;$SPM method}.
Our procedure for radius extraction is detailed in Ref.\,\cite{Cui:2021vgm}; so, we only recapitulate essentials.

The statistical SPM approach constructs a very large set of continued fraction interpolations, each element of which captures both local and global features of the curve the data are supposed to be measuring.  The global aspect is essential because it guarantees the validity of the interpolations outside the data range limits, ultimately enabling and justifying the evaluation of the curves' first derivative at the origin.

One must also account for the fact that good experimental data are statistically scattered around that curve which truly represents the observable whose measurement is the aim.  The data do not lie on the curve; hence, they should not be directly interpolated.  This issue can be addressed by \emph{smoothing} with a \emph{roughness penalty}, an approach we implement herein following the procedure in Ref.\,\cite{Reinsch:1967aa}.  A sketch of this method is provided in Ref.\,\cite[Sec.\,3]{Cui:2021vgm}.  It is characterised by a mathematically well-defined optimal roughness penalty, which is a self-consistent output of the smoothing procedure.  Denoting the roughness penalty by $\epsilon$, the data are untouched by smoothing if $\epsilon =0$; whereas maximal smoothing is indicated by $\epsilon =1$, which returns a linear least-squares realignment of the data.  In all pion cases described herein, $\epsilon \simeq 0$.

Becoming specific, the 1984 NA7 data set has $N=30$ elements on $0.015\leq Q^2/{\rm GeV}^2\leq 0.119$ \cite{Amendolia:1984nz}; and the 1986 NA7 set has $N=45$ elements on $0.015\leq Q^2/{\rm GeV}^2\leq 0.253$ \cite{Amendolia:1986wj}.  Preliminary SPM analysis of the latter data revealed that the $N=34$ lowest $Q^2$ points (square-within-square data in Fig.\,\ref{FAmendolia}) enable a robust radius determination.  Including the remaining points (half-shaded squares in Fig.\,\ref{FAmendolia}) serves only to introduce noise; so, we omit them from our detailed analysis.

To proceed with the statistical SPM method, we randomly select
$6 < M < N/2$
%
%
elements from each of the two NA7 data sets.  In principle, this provides $C(N,M)$ different interpolating functions in each case, amounting to ${\cal O}(10^5 - 10^7)$ or ${\cal O}(10^6 - 10^9)$ possible interpolators, respectively.
For each value of $M$, we then choose the first $5\,000$ curves that are smooth on the entire $Q^2$ domain covered by the data.  No further restriction is imposed; specifically, we do not require $F_\pi(Q^2=0)=1$.
Every interpolating function defines an extrapolation to $Q^2=0$, from which $r_\pi$ can be extracted using Eq.\,\eqref{Eqpirad}.  Thus, within a given data set, for each value of $M$, $r_\pi$ is determined by the average of all results obtained from the $5\,000$ curves.

In order to estimate the error associated with a given SPM-determined pion radius, one must first account for the experimental error in the data set.  We achieve this by using a statistical bootstrap procedure \cite{10.5555/1403886}, generating $1\,000$ replicas of the set by replacing each datum by a new point, randomly distributed around a mean defined by the datum itself with variance equal to its associated error.
%
Additionally, the fact that $M$ is not fixed leads to a second error source $\sigma_{{\delta\!M}}$, which we estimate by changing $M\to M'$, repeating the aforementioned procedure for this new $M'$-value, and evaluating the standard deviation of the distribution of $r^{M}_\pi$ for different $M$ values.

Consequently, the SPM result is
\begin{align}
	&r_\pi\pm\sigma_r;
	&r_\pi=\sum_{j=1}^{4}\frac{r^{{M_j}}_\pi}{4};&
&\sigma_r=\Bigg[\sum_{j=1}^{4}\frac{(\sigma^{{M_j}}_r)^2}{4^2}
+\sigma_{{\delta\!M}}^2\Bigg]^{\frac12}.
	\label{SPMrp}
\end{align}
Herein, for reasons explained below, we compute results for each one of the values $\{M_j=2+4j\,\vert\ j = 1,2,3,4\}$, so we have $20$--million 
results for $r_\pi$, each calculated from an independent interpolation.  Typically, $\sigma_{{\delta\!M}}\ll \sigma_{r}^{ M_j}$ for all $j$ values in the range specified above, i.e., the result is independent of our chosen value of $M$.

\medskip

\noindent\textbf{4.$\;$SPM validation}.
The first analysis stage is SPM validation; namely, establishing that the SPM radius extraction method is reliable when used in connection with both NA7 data sets.  To do this, we followed the procedure explained in Ref.\,\cite[Supplemental Material]{Cui:2021vgm}.

\emph{Step 1}.
Choose three specific functional forms that deliver a known value of the radius.  In this case, we used the radius in Eq.\,\eqref{PDFradius} and
\begin{equation}
\label{PiFF}
F_\pi(Q^2) \propto
\left\{
\begin{array}{lr}
\frac{1}{1 + r_\pi^2 Q^2/6} & \mbox{monopole} \\[1ex]
\frac{1}{(1 + r_\pi^2 Q^2/12)^2} & \mbox{dipole} \\[1ex]
{\rm e}^{-r_\pi^2 Q^2/6} & \mbox{Gaussian}
\end{array}\right.
\end{equation}
Working with these functions, we generated replica data sets built from values of the pion elastic electromagnetic form factor, $F_\pi$, evaluated at the $Q^2$ points sampled in each NA7 experiment.  The variability of real data was modelled by redistributing these values in response to fluctuations drawn according to a normal distribution.

\emph{Step 2}.
Treating each of the $3\times 2$ data sets obtained as real, we applied the SPM radius determination procedure described above.  Namely:
(\emph{i}) generate $10^3$ replicas;
(\emph{ii}) smooth each replica with the associated optimal parameter;
(\emph{iii}) use the SPM to obtain $r^{M_j}_\pi$ and $\sigma^{M_j}_\pi$, varying the number of input points $\{M_j=2+4j\,\vert\ j = 1,2,3,4\}$;
and
(\emph{iv}) calculate the final SPM result using Eq.\,\eqref{SPMrp} and compare it with the input value.

\underline{Remark~1}.
For a given value of $M\in {\cal S}_M=\{M_j=2+4j\,\vert\ j = 1,2,3,4\}$, all forms in Eq.\,\eqref{PiFF}, and each NA7 data set, the distribution of SPM-extracted radii is a Gaussian that is centred on the input value and possesses characteristics that are practically independent of $M$: in these cases, $\sigma_{{\delta\!M}}\ll \sigma^{{M_j}}_r$.

We also considered $M \in {\cal O}_M=\{ M_{ji} = 2 + 4 j + i\,\vert\, j=1,2,3,i=1,2,3\}$,
but in these instances one neither recovers the input radius nor obtains a Gaussian distribution.  Such exceptional $M$-values were not encountered when analysing the proton form factor data in Refs.\,\cite{Bernauer:2010wm, Xiong:2019umf}; so, their appearance is probably related to the character of the NA7 data.  Herein, therefore, we allow the validation procedure to impose a subset of $M$-values on our SPM analysis, \emph{viz}.\ $M\in {\cal S}_M$.

\begin{figure}[t]
\centerline{%
\includegraphics[clip, width=0.45\textwidth]{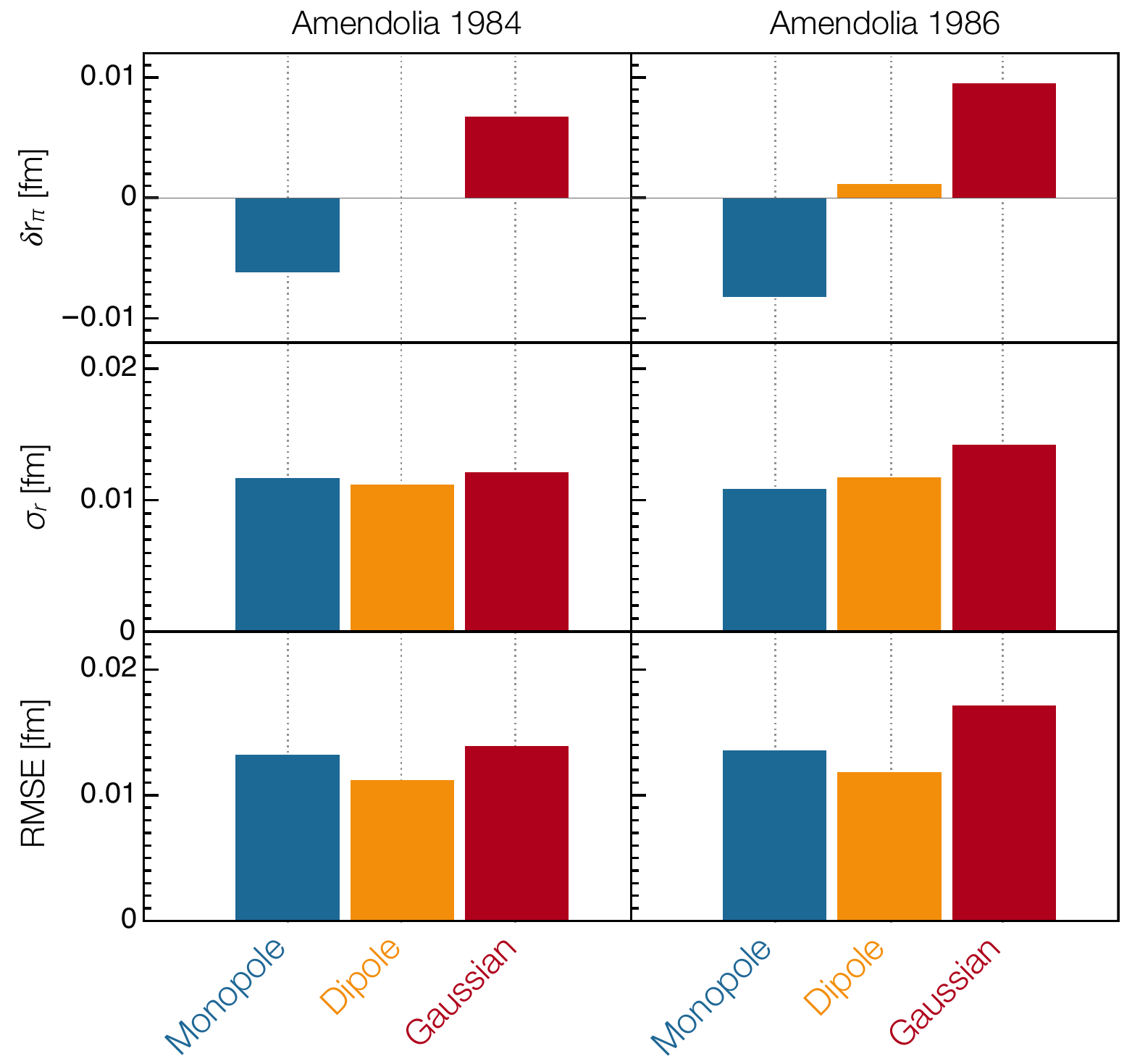}}
\caption{\label{FValidate}
Bias, $\delta r_\pi$, standard error, $\sigma_r$, and RMSE, Eq.\,\eqref{EqRMSE}, for SPM extrapolations of the pion radius from $10^3$ replicas generated using the three models in Eq.\,\eqref{PiFF}.  Notably, the RMSE is approximately independent of the model used.
\emph{Left panels} -- data from Ref.\,\cite{Amendolia:1984nz}; and
\emph{right panels} -- $N=35$ data in Fig.\,\ref{FAmendolia}, drawn from Ref.\,\cite{Amendolia:1986wj}.
}
\end{figure}

\underline{Remark~2}.
Defining the bias as \mbox{$\delta r_\pi=r_\pi-r^{\rm input}_p$}, then the SPM extraction of the pion radius is robust: in all cases, $|\delta r_\pi| \lesssim \sigma_r$, where $\sigma_r$ is the standard error in Eq.\,\eqref{SPMrp}.
This is demonstrated by Fig.\,\ref{FValidate}, which displays $\delta r_\pi$ as obtained using the SPM to extract $r_\pi$ from the three generators in Eq.\,\eqref{PiFF}.

\underline{Remark~3}.
Defining the root mean square error (RMSE)
\begin{equation}
	\mathrm{RMSE}=\sqrt{(\delta r_\pi)^2+\sigma_r^2}\,,
\label{EqRMSE}
\end{equation}
then, as shown in Fig.\,\ref{FValidate}, the SPM analysis produces RMSE values that are approximatively independent of the model used to obtain the replicas.  Hence, the SPM procedure satisfies a standard ``goodness of fit'' criterion \cite{Yan:2018bez}, in consequence of which our SPM extractions of $r_\pi$ can objectively be judged as a sound expression of the information contained in the data.

\medskip

\noindent\textbf{5.$\;$Pion charge radius}.
Having validated the SPM procedure, we turn now to the real data.  First, however, we list the results originally presented.
The 1984 experiment \cite{Amendolia:1984nz} reported charge radii obtained from their data using the following $F_\pi$ fit forms: monopole -- $r_\pi = 0.657 \pm 0.012\,$fm; dipole -- $r_\pi = 0.641 \pm 0.012\,$fm; and a sum of orthogonal polynomials, reaching to O$(Q^4)$ -- $r_\pi = 0.660 \pm 0.024\,$fm.
Turning to 1986, Ref.\,\cite{Amendolia:1986wj} reported $r_\pi = 0.653 \pm 0.008\,$fm from their data; and incorporating additional information from $\pi\pi$ partial wave analyses, $r_\pi = 0.663 \pm 0.006\,$fm.
All originally reported values are mutually consistent within quoted uncertainties; but some sensitivity to the fitting assumptions is evident in both the central values and uncertainties.

\begin{figure}[t]
\centerline{%
\includegraphics[clip, width=0.42\textwidth]{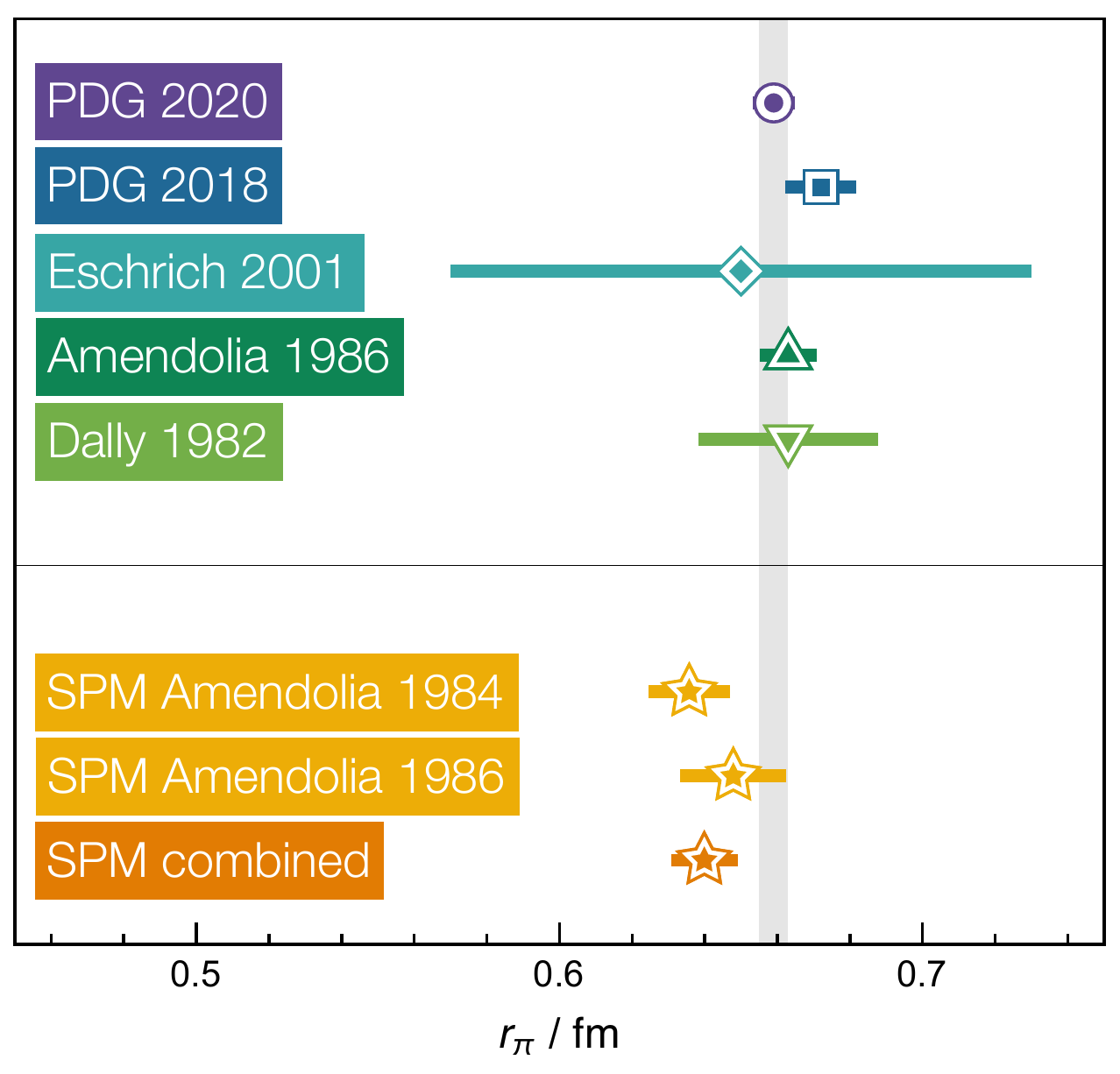}}
\caption{\label{Fradpi}
Selected published values of the pion charge radius:
purple circle (Particle Data Group -- PDG, 2020) \cite{Zyla:2020zbs};
blue square (PDG 2018) \cite{Tanabashi:2018oca};
cyan diamond \cite{GoughEschrich:2001ji};
green up-triangle \cite{Amendolia:1986wj};
light-green down-triangle \cite{Dally:1982zk}.
The vertical grey band delimits the uncertainty in Eq.\,\eqref{PDFradius}.
Results obtained herein -- gold stars:
SPM A-84 from Ref.\,\cite{Amendolia:1984nz};
SPM A-86 from Ref.\,\cite{Amendolia:1986wj}; and orange star -- SPM combined, the average in Eq.\,\eqref{rpiNA7}.
}
\end{figure}

As emphasised above, the SPM returns a function-form-unbiased result; and working with the 1984 NA7 data \cite{Amendolia:1984nz}, the SPM yields
\begin{equation}
\label{rpi84}
\,_{\rm SPM}r_\pi^{\rm 84} = 0.636 \pm 0.009_{\rm stat}\;{\rm fm}\,.
\end{equation}
Considering instead the $35$ low-$Q^2$ points in Fig.\,\ref{FAmendolia}\,--\,lower panel, we obtain
\begin{equation}
\label{rpi86}
\,_{\rm SPM}r_\pi^{\rm 86} = 0.648 \pm 0.013_{\rm stat}\;{\rm fm}\,,
\end{equation}
which is consistent with Eq.\,\eqref{rpi84}.  In each case, the SPM returns a value $\approx 1\,\sigma$ below the error-weighted average of the original function-choice dependent determinations: $\bar r_\pi^{84}=0.650(8)\,$fm and $\bar r_\pi^{86}=0.659(5)\,$fm, respectively.

The results in Eqs.\,\eqref{rpi84}, \eqref{rpi86} are compared with other determinations in Fig.\,\ref{Fradpi}.  Notably, our reanalysis of the NA7 data sets supports the recent downward shift of the PDG average.

\medskip

\noindent\textbf{6.$\;$Summary, kaon radius, and perspectives}.
Using a statistical sampling approach based on the Schlessinger Point Method for the interpolation and extrapolation of smooth functions, which is practically equivalent to the multipoint Pad\'e approximants technique, we extracted the pion charge radius, $r_\pi$, by analysing the highest-precision, lowest-$Q^2$ $\pi e$ elastic scattering data that is currently available \cite{Amendolia:1986wj, Amendolia:1984nz}.  A key feature of our scheme is that no specific function form is assumed for the interpolator; hence, it supplies form-unbiased interpolations as the basis for well-constrained extrapolations.  Consistent results were obtained from the two data sets [Eqs.\,\eqref{rpi84} and \eqref{rpi86}],
the error-weighted average of which is
\begin{equation}
\label{rpiNA7}
\,_{\rm SPM}r_\pi^{\rm NA7} = 0.640 \pm 0.007_{\rm stat}\;{\rm fm}\,.
\end{equation}

Available $\pi e$ elastic scattering data do not match analogous modern electron+proton data in precision, density of coverage, or low-$Q^2$ reach.  Given the importance of all these things in resolving the so-called proton radius puzzle, it would be reasonable to withhold judgement on the true value of $r_\pi$.  New precise $\pi e$ data, reaching as far as possible below $Q^2 = 0.015\,$GeV$^2$ and densely covering the associated low-$Q^2$ domain, are desirable.

\begin{figure}[t]
\centerline{%
\includegraphics[clip, width=0.45\textwidth]{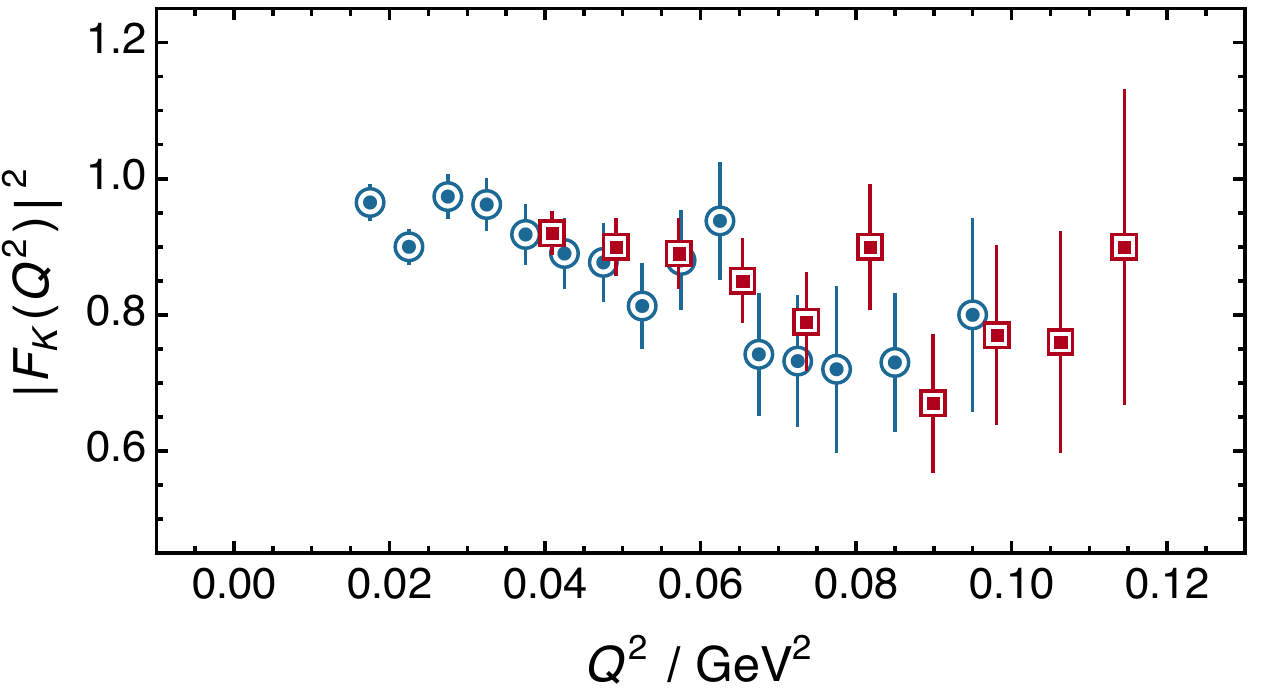}}
\caption{\label{FKaon}
Available low-$Q^2$ kaon form factor data (red squares \cite{Dally:1980dj} and blue circles \cite{Amendolia:1986ui}) used as the basis for existing extractions of the charged-kaon electric radius \cite{Zyla:2020zbs}.
}
\end{figure}

In the absence of Higgs couplings into QCD, the kaon would be indistinguishable from the pion and, hence, also a Nambu-Goldstone boson.  Differences between the $\pi$ and $K$ express Higgs-boson modulations of EHM, i.e., interference effects between Nature's two known sources of mass; and one of the simplest examples is the ratio of kaon and pion charge radii.  However, regarding $r_K$, the charged-kaon radius, the situation is far worse than for the $\pi$.  This is apparent in Fig.\,\ref{FKaon}, which reproduces the data used for extraction of $r_K$ \cite{Dally:1980dj, Amendolia:1986ui}.  Such data is too imprecise for any reasonable determination of $r_K$.  The commonly quoted value \cite{Zyla:2020zbs}:
\begin{equation}
\label{kaonradius}
r_K = 0.560 \pm 0.031\,{\rm fm}\,,
\end{equation}
is obtained by averaging the results of separate monopole-squared fits to the distinct data sets in Refs.\,\cite{Dally:1980dj, Amendolia:1986ui}.  Exploiting Eq.\,\eqref{Eqfpirpi}, then Eq.\,\eqref{PDFradius} entails $r_K = 0.552(3)\,$fm.

Employing the SPM, we find when smoothing that the largest possible roughness penalty is returned, $\epsilon = 1$.  Mathematically, this means that the information contained in the data is insufficient to yield an objective result for the radius: the value obtained is strongly influenced by the assumed fitting form.  With $\epsilon = 1$, the SPM interpolants relax to a linear least-squares fit to the data in Fig.\,\ref{FKaon}, which returns $\chi^2/{\rm datum} = 0.68$ and
\begin{equation}
r_K = 0.53\,{\rm fm}\,.
\end{equation}
This value is consistent with Eq.\eqref{kaonradius}, an outcome which further highlights the inadequacy of available data so far as a precise determination of the kaon radius is concerned.  (Eq.\,\eqref{Eqfpirpi} produces $r_K=0.536(6)\,$fm from Eq.\,\eqref{rpiNA7}.)

High-luminosity $\pi$ and $K$ beams are planned in connection with a new QCD facility at the CERN SPS \cite{Adams:2018pwt, Andrieux:2020}; so, one can expect the coming decade to deliver the first new $\pi e$ and $K e$ elastic scattering data in roughly forty years.  Should such data meet the demands of precision, low-$Q^2$ reach, and density of coverage that are required, then they will enable objective determinations of pion and kaon charge radii.

\medskip
\noindent\emph{Acknowledgments}.
We are grateful for constructive comments from O.~Denisov, J.~Friedrich, W.-D.~Nowak and C.~Quintans.
Use of the computer clusters at the Nanjing University Institute for Nonperturbative Physics is gratefully acknowledged.
Work supported by:
National Natural Science Foundation of China (under Grant No.\,11805097);
and Jiangsu Provincial Natural Science Foundation of China (under Grant No.\,BK20180323).
%
%



\end{document}